\documentclass[12pt,preprint]{aastex}

\slugcomment{\today}

\def\los{{\it l.o.s.\,}}
\def\ie{{\it i.e.\,}}
\def\ea{{\it et al., \,}}
\def\be{\begin{equation}}
\def\ee{\end{equation}}

\begin{document}
\title{Millimeter observation of the SZ effect in the Corona Borealis supercluster}

\author{E.S.~Battistelli$^{1,2}$\altaffilmark{,6}, M.~De~Petris$^{2}$,
L.~Lamagna$^{2}$, R.A.~Watson$^{3,7}$, R.~Rebolo$^{1,4}$,
F.~Melchiorri$^{2}$, R.~G\'enova-Santos$^{1}$, G.~Luzzi$^{2}$,
S.~De~Gregori$^{2}$, J.A.~Rubi\~no-Martin$^{1}$, R.D.~Davies$^{3}$,
R.J.~Davis$^{3}$, K.~Grainge$^{5}$, M.P.~Hobson$^{5}$,
R.D.E.~Saunders$^{5}$, and P.F.~Scott$^{5}$}

\affil{$^{1}$Instituto de Astrof\'isica de Canarias, C/ Via Lactea
s/n, 38200, La Laguna, Spain}

\affil{$^{2}$Department of Physics, University "La Sapienza", P.le
A. Moro 2, 00185, Rome, Italy}

\affil{$^{3}$Jodrell Bank Observatory, University of Manchester,
Macclesfield, Cheshire, SK11 9DL, UK}

\affil{$^{4}$Consejo Superior de Investigaciones Cient\'ificas,
Spain}

\affil{$^{5}$Astrophysics Group, Cavendish Laboratory, University of
Cambridge, Cambridge, CB3 OHE, UK}

\altaffiltext{6}{current address: University of British Columbia,
Department of Physics and Astronomy, 6224 Agricultural Road,
Vancouver, B.C. Canada V6T 1Z1 \\ e-mail:
elia.stefano.battistelli@roma1.infn.it}

\altaffiltext{7}{current address: Instituto de Astrof\'isica de
Canarias, C/ Via Lactea s/n, 38200, La Laguna, Spain}

\begin{abstract}

We have observed the Corona Borealis supercluster with the
Millimeter and Infrared Testa Grigia Observatory (MITO), located in
the Italian Alps, at 143, 214, 272, and 353 GHz. We present a
description of the measurements, data analysis, and results of the
observations together with a comparison with observations performed
at 33 GHz with the Very Small Array (VSA) interferometer situated at
the Teide Observatory (Tenerife, Spain). Observations have been made
in the direction of the supercluster towards a cosmic microwave
background (CMB) cold spot previously detected in a VSA temperature
map. Observational strategy and data analysis are described in
detail, explaining the procedures used to disentangle primary and
secondary anisotropies in the resulting maps. From a first level of
data analysis we find evidence in MITO data of primary anisotropy
but still with room for the presence of secondary anisotropy,
especially when VSA results are included. With a second level of
data analysis using map making and the maximum entropy method we
claim a weak detection of a faint signal compatible with a SZ effect
characterized at most by a Comptonization parameter
$y=(7.8^{+5.3}_{-4.4})\times10^{-6}$ 68\% CL. The low level of
confidence in the presence of a SZ signal invite us to study this
sky region with higher sensitivity and angular resolution
experiments such as the already-planned upgraded versions of VSA and
MITO.

\end{abstract}


\section{Introduction}\label{par:intro}

The cosmic microwave background radiation (CMB) is one of the most
powerful tools in cosmology. Distortions of the CMB frequency
spectrum due to relic photon inverse Compton scattering with hot
electron gas in clusters of galaxies is called the
Sunyaev-Zel'dovich effect (SZE) \cite{SunZel72}; reviews of the
effect can be found in Rephaeli (1995), Birkinshaw (1999), and
Carlstrom \ea (2002). This secondary anisotropy effect produces a
brightness change in the CMB that can be detected both at radio
frequencies and at millimeter and sub-millimeter wavelengths,
appearing as a negative signal (with respect to the average CMB
temperature) at frequencies lower than $\sim$217 GHz and as a
positive signal at higher frequencies. Its relatively simple
spectral behavior and the fact that in the Cosmological Standard
Model it is redshift independent, make the SZE a powerful tool, both
for investigating the physical properties of the cluster of galaxies
through which it is observed and for extracting important
cosmological information like the Hubble constant or the absolute
temperature of the CMB at the redshift of the cluster of galaxies.

The SZE intensity change depends on the electron density of the
scattering medium $n_{e}$ and on the electron temperature $T_{e}$
both integrated over the line of sight (\los) $l$, and can be
described by the Comptonization parameter $y$: \be y=\int
n_{e}\sigma_{T}\frac{k_{B}T_{e}}{m_{e}c^{2}} dl,\ee where
$\sigma_{T}$ is the Thomson cross section, $k_{B}$ is the Boltzman
constant, $m_{e}$ is the electron mass, and $c$ is the light speed
in vacuum. In a non-relativistic approximation, the thermal SZE
spectral behavior is described, in terms of the thermodynamic
temperature change $\Delta T_{SZ}$ with respect to the average CMB
temperature $T_{CMB}$, by the expression: \be \frac{\Delta
T_{SZ}(x)}{T_{CMB}}=y(x\cdot coth(x/2)-4),\ee where $x$ is the
adimensional frequency $x=h\nu /(k_{B}T_{CMB})$.

So far, the SZE has only been observed in the direction of clusters
of galaxies; however, other objects may also be sources of
detectable SZE. Clusters of galaxies are often found in
superclusters of galaxies (SCG). The intra-supercluster medium
(ISCM) present in their central regions, despite the relatively low
density and temperature of the baryon population, may be
sufficiently elongated along the \los to be detectable through SZE
even without galaxy clusters being present throughout the same
region.

The baryon distribution in the Universe is in fact still an open
problem for modern cosmology. The observed baryonic matter in the
local Universe obtained from H I absorption, gas and stars in galaxy
clusters, and X-ray emission is still small compared to what is
predicted by nucleosynthesis (see e.g. Fukugita \ea 1998), by
Ly$\alpha$ forest absorption observations (see e.g. Rauch \ea 1997),
and by measurements of the CMB power spectrum (see e.g. Spergel \ea
2003). A diffuse baryonic dark matter detection could explain, at
least in part, the apparent discrepancy between the observed and the
expected baryon density. Simulations of large-scale cosmological
hydrodynamic galaxy formation (see e.g. Cen and Ostriker, 1999) show
that at low redshift, these missing baryons, which represent
approximately half of the total baryon content of the Universe,
should lie in the temperature range of $10^{5} \rm K<T<10^{7}\rm K$
in a state of warm-hot gas not yet observed through their EUV and
soft-X-ray emission. Yoshida {\it et al.} (2005) have studied the
temperature structure of this warm-hot intergalactic medium (WHIM).

Hence, superclusters seem to be attractive regions in which to
search for these missing baryons. Indeed, many authors have
conducted X-ray measurements of their emission (e.g. Persic \ea
1988, 1990; Day \ea 1991; Bardelli \ea 1996; Zappacosta \ea 2005
Rephaeli and Persic 1992), but finding no compelling evidence for
the presence of WHIM in the observed regions. Inside superclusters
of galaxies, the outskirts of clusters of galaxies have been studied
by Finoguenov (2003), Kaastra \ea (2003), and Bonamente \ea (2003),
who found evidence for soft X-ray excess, which could be due to warm
baryonic diffuse matter. These are of fundamental importance in
understanding the structure of the superclusters themselves.

In addition to the X-rays, the SZE may represent a useful tool to
search for the WHIM because, as stated above, the large path lengths
of the CMB photons across SCGs could produce a significant effect.
At angular scales of a few square degrees (typical size of the known
SCGs), limits on the diffuse SZE have been placed by Banday {\it et
al.} (1996) and Rubi\~no-Mart\'{\i}n {\it et al.} (2000), who cross
correlated the COBE DMR all-sky data and the Tenerife data,
respectively, with catalogues of clusters of galaxies finding a
stringent upper limit of $\approx 8 \mu \rm K$ at, respectively,
$7^{\circ}$ and $5^{\circ}$ angular scales. However, at smaller
angular scales, a deep study of this diffuse emission is only now
becoming effective, with the increase in the number of experiments
whose primary goal is to map the microwave sky both characterizing
primary and secondary CMB anisotropies and extracting important
cosmological information from these maps. Statistical work has, for
instance, been carried out using WMAP first-year data with given
catalogs of galaxy clusters (Hern\'andez-Monteagudo \ea 2004a,
2004b; Myers \ea 2004; Lieu \ea 2005).

In section \ref{par:VSA} we briefly report the VSA observation of
the Corona Borealis supercluster, and in section \ref{par:obs} the
MITO experiment is introduced and the observational strategy is
presented. In section \ref{par:sim} we present the use of
simulations in the data analysis stage. In sections \ref{par:anal}
and \ref{par:cont} data analysis is described with the discussion of
the possible sources of contamination. Finally in section
\ref{par:res} we present and discuss the results.

\section{VSA observation on the Corona Borealis supercluster}\label{par:VSA}

The VSA instrument is a 14 element heterodyne interferometer array
with 1.5 GHz bandwidth, tunable between 26 and 36 GHz, situated on
the Teide Observatory in Tenerife (see e.g. Watson \ea 2003). It is
the result of an Anglo-Spanish collaboration between the Cavendish
Laboratory, the Jodrell Bank Observatory, and the Instituto de
Astrof\'isica de Canarias. In Extended Configuration it uses
identical conical corrugated horns 322 mm in diameter with a primary
beam of $2^{\circ}.1$ FWHM and a synthesized beam of $\approx 11'$
FWHM at 33 GHz.

The Corona Borealis supercluster has been mapped by VSA mainly in
the period between the end of 2003 and the middle of 2004
\cite{gen05a,gen05b}. A major result appearing in these maps is the
presence of two strong ($\approx 8 \sigma$) temperature decrements
near the center of the SCG in directions with no known cluster of
galaxies with minimum intensities of -72$\pm$12 and -103$\pm$10
mJy/beam respectively, with typical sizes of $\approx$3 VSA
synthesized beams (thermodynamic temperatures correspond
respectively to -157$\pm$27 and -230$\pm$23 $\mu$K). The two spots
(namely spot B and H in the pointing mosaic carried out by VSA) are
centered on: R.A.=$15^{h} 25^{m} 21^{s}.60$ dec=$+29^{\circ} 32'
40''.7$ and R.A.=$15^{h} 22^{m} 11^{s}.47$ dec=$+28^{\circ} 54'
06''.2$ (J2000). Monte Carlo simulations have highlighted the
unlikeliness of these spots (especially the strongest one) to be due
only to primary CMB anisotropies with a probability 37.82$\%$ and
0.38$\%$ respectively. Press-Schechter and Sheth-Tormen mass
functions were used in order to investigate the possibility that
these anisotropies could be due to unresolved and unknown cluster of
galaxies, showing again a low possibility for this to happen. The
probability of these spots arising from SZE, and in particular from
extended SZE, is thus what we focus on in this paper.

The evidence of these SCG spots being due to inverse Compton
scattering would result from a positive signal at frequencies higher
than $\approx$217 GHz, an effect that uniquely characterizes the
SZE. Observations have thus been performed at MITO (with four
channels centered at 143, 214, 272, and 353 GHz), which is capable
of disentangling sources of anisotropy with different spectral
behavior.

\section{MITO observations}\label{par:obs}

\subsection{The instrument}

The MITO experiment is a 2.6-m primary mirror telescope equipped
with a four channel photometer (FotoMITO) operating from Italian
Alps at 3500 m altitude \cite{depe99,depe05a}. It is a Cassegrain
telescope in altitude-azimuthal configuration optimized for
differential measurements, thanks to the secondary mirror, which can
wobble around the neutral point with digitally controlled wave
forms. FotoMITO is a single pixel, multifrequency photometer with
four neutron transmutation-doped (NTD) Ge composite bolometers
cooled to 290 mK. The 4 K cooled refocusing optics, mesh filters,
and beam splitters allow a reduction of the background on the
bolometer as well as the selection of the desired frequency
bandwidth and throughput, together with the 290 mK cooled Winston
cones working as radiation collectors. The working frequencies,
summarized in table \ref{tab:char}, have been selected to perform
efficient SZE studies and to match the 2 mm, 1 mm, and 850 $\mu$m
atmospheric windows. The beam of the instrument has a FWHM of $16'$
\cite{sav03} and the pointing accuracy is $\approx 1'$ achieved by
frequent observation of bright stars close to the source with a CCD
camera. The MITO experiment has been efficiently used to perform
measurements of the SZE from the Coma Cluster (De Petris \ea 2002;
Savini \ea 2003) as well as to extract important cosmological
information from it \cite{bat02,bat03}. Table \ref{tab:char} gives
the main MITO characteristics.

\subsection{The observational strategy}

Data were collected in 2004 March and include observations on the
deepest spot of the VSA map (\ie spot H), calibration scans on
Jupiter, some pointings on Saturn and Tau A used as secondary
calibration sources, and routine sky dips to extract information
about atmospheric emission.

Sky modulation has been performed by a three fields square-wave-like
scanning along constant elevation followed by lock-in demodulation.
This allows efficient removal of the atmosphere emission even when
it exhibits strong linear gradients parallel to the horizon. The
beamthrow was set to $41'$ and the modulation frequency to 4.5 Hz
with second harmonic demodulation.

Measurements were carried out by performing drift scans (DSs) on the
source. This procedure, with respect to antenna nodding (AN)
procedure, allows the reduction of bolometer microphonics as the
telescope is at rest during the observation. It also helps
separation of side-lobe pick-up from sky signal and permits an
efficient monitoring of the atmosphere in the "off-region" position
before and after the source crosses the instrument beam; in general,
it allows substantial control of many systematics. On the other
hand, in scanning mode the effective integration time on the source
is reduced, and due to the source motion in the sky, a rotation of
the modulation reference plane occurs between various DSs, resulting
in a more complicated procedure for scan combination (this problem
is also present for AN observations).

We have collected 105 scans across the region of Corona Borealis at
the VSA H spot. Each scan is 10 minutes long in R.A.

\section{Simulations}\label{par:sim}

Given the adopted scan-mode strategy and the differential
measurements, a fitting of the spatial distribution of the observed
signals requires information about the instrument (beam aperture and
modulation strategy) and about the expected brightness profile of
the source. Unfortunately, in the present case, this procedure
introduces a higher uncertainty with respect to similar ones already
used for SZE observations of galaxy clusters \cite{depe02} whose
characteristics are well constrained by X-ray observation.

Simulations have thus been performed from the CLEANed VSA Corona
Borealis map and compared with real scans. CMB primary anisotropy
maps allowing for the presence of secondary anisotropy signal with
different shape and intensity have in fact been produced at MITO
frequencies convolving the signal with the channels' response. The
maps have then been normalized to the signal we get in the VSA map.
Analogue procedures have then been applied to the $13'$ FWHM WMAP
90-GHz map with no significant detection mainly due to the low S/N
of the first year data in the specific location.

DS procedures have thus been simulated on the computed maps at the
experimental coordinates where effective measurements have been
carried out. The scan simulation has been performed accounting for
the MITO beam size as well as for the bolometer time constants and
the lock-in transfer function. Like the real MITO data, simulated
scans with the MITO strategy appear as four time ordered data
sequences.

The map normalization and the scan simulation procedure
automatically account for the effect of beam dilution and reference
beam contamination often considered as a later correction included
with the form factor. The comparison between the simulations and the
real scans accounts for the effect of the MITO window function that
would drive a sky measurement away from the real sky temperature
because of dilution and field contamination.

Further simulations have been performed on maps calculated from the
best-fit analysis in order to determine the most probable maps
consistent with MITO data. This method is presented in section
\ref{iter}.

\section{Data analysis} \label{par:anal}

\subsection{Preliminary procedures} \label{analpre}

Our DSs are characterized by offsets due to several sources that
create modulated signals difficult to extract from data mainly due
to small asymmetries. Offset sources have been studied and simulated
and much effort has been spent in order to reduce and stabilize
those. However, a small source of offset is present in our data
which we can attempt to remove at the data analysis stage through a
constant or linear fitting of it. A quadratic fitting did not change
the $rms$ value of the residual fluctuation, although it might
affect the window function at low multipoles.

Another feature evident in the MITO data is the presence of spikes
due to both cosmic rays heating our bolometers and electromagnetic
interference. These spikes are simply excluded with a dedicated
algorithm.

\subsection{Atmospheric transmission}

As is well known, even when observing from sites at high altitude,
atmospheric transmission at millimeter and submillimeter wavelengths
is greatly affected by precipitable water vapor (PWV) content.
Measurements of atmospheric emission have thus been taken and
compared with models describing atmospheric characteristics in order
to perform estimates of its transmission.

Atmospheric emission has been monitored by performing sky dips, \ie
measuring atmospheric emission at different elevations in order to
test its dependence on the air mass. Measurements have been
performed by focal plane chopping between atmospheric signal and a
black body reference (\ie Eccosorb {\it AN-72}) at ambient
temperature.

We have used the Atmospheric Transmission at Microwaves (ATM)
program \cite{par01} to simulate the atmospheric emission, at
different elevations, PWV, atmospheric pressure, and temperature.
Uncertainties due to the responsivity determination have been
minimized by calculating the ratios between signals at different
elevation. Experimental data have then been fitted to the models
estimating the value of the PWV for each observing night. In this
way we have estimated atmospheric transmission for the four MITO
channels during each scan of the observed region. In table
\ref{tab:char} we report the average transmissions $<\tau>$ for the
four MITO channels.

In order to perform the best atmosphere transmission monitoring
simultaneous with the observations, a dedicated spectrometer is
under development \cite{depe05b}.

\subsection{Calibration}

Calibrations are performed observing planet sources such as Jupiter
which is one of the brightest "point-source" at millimeter and
sub-millimeter wavelengths. Jupiter has also been used for a precise
measurement of the effective beam shape \cite{sav03}. The planet
emission at millimeter and sub-millimeter wavelengths has been
measured and simulated through many models (we used Moreno, 1998).
The lack of information about its emission process results in a
typical uncertainty of 10$\%$ on the temperature value. For the four
MITO channels we have 171, 171, 170, and 169 K at 143, 213, 272, and
353 GHz, respectively.

From the Jupiter observations, we extract the calibration factors
$C$ whose averages are summarized in table \ref{tab:char}. DSs have
also been performed at different declinations and on different
sources such as Saturn and Tau A as a consistency check, to map the
sources and to check for bad alignments. The errors are dominated by
the 10\% uncertainties over the planet absolute temperature. Small
differences in $C$ for different nights of observation have been
encountered and have been found to be due to the intrinsic drifts in
the receiver response and instrumental setups; we have accounted for
this by calibrating our signals night by night.

Figure \ref{jup} shows the signals detected in the four MITO
channels in one of the Jupiter scans.

\subsection{Noise filtering and atmospheric decorrelation}\label{par:noise}

As already stated, three field chopping measurements are
intrinsically insensitive to offset and linear gradients in the
atmospheric emission. However, atmospheric temporal and spatial
fluctuations are still present in our data, dominating the
cosmological signal of interest. In order to get rid of this
contamination, we have applied an atmospheric decorrelation based on
the large overlapping of the four MITO channel beams (\ie deviations
are lower than 1$\%$) and the presence of the 353-GHz channel, which
is much more sensitive to the atmospheric emission. The basic
procedure is to subtract, from the channels at the frequency of
cosmological interest, the channel most affected by atmospheric
emission, best fitting the factor driving the latter into the other
channels. High correlation between the channels is a key ingredient
necessary to performing an efficient decorrelation. Uncorrelated
signals may be due to different optical depths for different
spectral channels, to fluctuations at different altitudes, or to
differences in the channel beam (for a detailed discussion on this
topic see Melchiorri and Olivo-Melchiorri, 2005).

Analysis and simulations are being carried out in this field and
will be described elsewhere (S. De Gregori \ea 2006, in
preparation). The key point is that under average PWV conditions
(\ie $\approx$1mm at Testa Grigia Observatory) and for the MITO
spectral bandwidths, the correlation factors between the 353 GHz
channel and the others show a high stability against water vapor
change for the 143 GHz channel, and with moderate common variation
when dealing with the 214 and 272 GHz channels. As a result, the
decorrelation operated from the high frequency channel gives good
efficiency when applied to the 143 GHz channel while leaving
residuals in the other two channels depending on the amount of water
vapor change within the duration of the scan. Because of the
nonlinear dependence with the PVW of the atmospheric signal in the
high frequency channels, these residuals may be characterized by a
non-zero average and leave an imprint in our decorrelated signals.

Many scans collected in this observing campaign have shown high
enough correlation to allow an efficient subtraction (threshold has
been set at 85$\%$); however, in some cases the correlation has not
been found high enough to allow the decorrelation procedure to
approach the detector noise. In these cases, further filtering
procedures have been applied in order to suppress the uncorrelated
components in the signals. We have tested different filtering
procedures in both real space and Fourier space. These procedures
have been found efficient for those scans showing moderately high
correlation even if not enough to undergo an efficient
decorrelation.

The filtering procedure that we finally implemented is presented in
Melchiorri and Olivo-Melchiorri (2005) and consists of passing data
through a non linear amplifier that yields an output signal
$S_{out}$ related with the input signal $S_{in}$ by \be
S_{out}=S_{in}exp[-\frac{|S_{in}|}{\Gamma}], \label{eqfilog} \ee
with $\Gamma \sim 1\sigma$ to be found from simulations considering
the effective uncorrelated noise present in out data. This technique
has been applied on the ratios between the various channels. The
filter has been calibrated from S/N considerations and has been
found efficient in removing non-Gaussian noise components, allowing
us to approach the fundamental noise of the detectors. Table
\ref{tab:decor} shows the $rms$ of the signals of the first three
MITO channels before and after the filtering procedure. The
systematics introduced by this procedure have been simulated and
found to be negligible compared to the random errors we have at the
end of the procedure. The non linear amplifier allows us to "clean"
our data of non-Gaussianity in an unbiased way, controlling the
systematics.

About 1/3 of the total number of scans have been removed from the
analysis for not showing high enough correlation to be efficiently
decorrelated. The level of decorrelation has been chosen from
simulations. The main average characteristics of our filtered and
decorrelated scans are reported in table \ref{tab:decor}. We have
reported the average correlation factors for the first three
channels with the 353 GHz channel before and after the filtering
procedure, color ratios with respect to the 353-GHz channel together
with the dispersion of the other channels, the $rms$ before and
after the filtering procedure and the decorrelation.

Once decorrelated, the scans have been found to be Gaussian
distributed around their average values. In this case, the impact on
the results has also been found to be negligible by applying the
process to simulated data sets.

\subsection{Best fit}\label{par best fit}

We have applied different methods to extract the information of
interest from our cleaned and decorrelated scans. They give
consistent results, and meaningful information has finally been
extracted by applying maximum entropy method (MEM) to produce the
final maps, Monte Carlo (MC) simulations for estimating the errors
and maximum likelihood analysis to determine the effective
contribution of primary and secondary anisotropies in our maps.

As a first attempt we have applied the method already used for SZE
extraction from galaxy cluster observation performed by MITO
\cite{depe02}. In this case, the best fitting of the atmospheric
correlation factors is performed together with the best fitting of
the parameters describing the intensity of the observed cosmological
signal. The latter factors are in fact multiplied by the simulations
presented in section \ref{par:sim}, which describe the expected
shape of the detected signal. For each scan, we have thus performed
a maximum likelihood analysis and the results have been combined to
extract the final result. Different hypotheses have been tested by
using different simulations. This procedure introduces, in the
present case, a higher uncertainty with respect to galaxy cluster
observations that take advantage of detailed physical information
arising from X-ray observations. As a result, the combination of the
fits does not allow us to extract meaningful conclusions due to the
lack of information about the physical properties of the diffuse gas
present in the observed region.

An alternative approach consists of performing the decorrelation
without subtracting the simulation from the 353 GHz channel. This
procedure introduces a systematic error due to the presence, in the
high frequency channel, of a cosmological signal. However, this
decorrelation procedure has the advantage of being totally
independent of the assumption about the cosmological signal we are
searching for. Moreover, due to the large atmospheric signal present
in the 353 GHz channel, this systematic error can be kept under
control. In table \ref{tab:decor} we list the average
proportionality factors between the atmospheric noise in the various
channels with the 353 GHz channel. The systematics introduced in
this way, related to a primary anisotropy detection, contribute
factors of around 9\%, 7\%, and 14\% respectively in the first three
MITO channels, while for SZE detections they contribute 19\%, 129\%,
and 32\% respectively considering the spectral signature of the
effect. These corrections are taken into consideration in the
subsequent analysis.

The first attempt to use the subsequent scan is by simply averaging
them and comparing with the (averaged) simulations performed on the
VSA map. It is worth stressing that these averages do not correspond
to what is observed in the sky due to the previously mentioned
rotation of the reference fields. However, this analysis allows us
to extract information on the observed signal from its spectral
behavior. From this comparison it is clear that a primary anisotropy
signal is present in our data, but still leaving room for further
secondary effect like SZE \cite{bat05}. However, the combination of
our data and simulations do not allow us to $spectrally$
characterize SZE signals present in our data.

\subsubsection{Iterative fitting method}\label{iter}

We have developed an analysis method capable of generating maps from
our differential measurements. Initially this has been performed
with an iterative method; starting from the 33 GHz VSA map, we have
corrected, at the declination of the central beam of the MITO scans,
for the difference between the measured scans and the DS simulation
performed on the VSA map. We have then performed simulations on the
corrected maps, and we have repeated the procedure until
convergence. In this way we obviously introduce a prior based on the
assumption that the lateral fields of the MITO scans are those
reported in the VSA map. This procedure is based on the hypothesis
that the average of the lateral fields in the new MITO corrected
maps does not change during the iteration. This is not exactly true,
even if this effect has been shown to be small by performing several
iterations of the map-making procedure. This procedure is (still)
not independent of the VSA observations, but the assumption about
the reference MITO fields is weaker than in the method described
above. We conclude there is a signal in the H spot that is dominated
by a primary anisotropy but with an additional signal with a
frequency dependence characteristic of a rising spectrum signal (as
the SZ effect is in our spectral range). Still, from simulations we
felt that the degeneracy between different realizations of the sky
resulting in the same scans patterns is not totally controlled. For
this reason we developed further map-making analysis in order to
further probe the contributions.

\subsubsection{Maximum Entropy method}\label{MEM}

As map reconstruction is degenerate for differential and
interferometer observations, an extra regularizing constraint is
required which is conveniently provided by MEM. In order to extract
the best representative signal maps from the original data sets, we
have carried out the MEM reconstruction for both the MITO
decorrelated scans and the gridded VSA visibilities for point H in
the Corona Borealis supercluster. We used the same method of fitting
modeled data resulting from a trial sky for both data sets in order
to be sure of consistency. The sky model used comprises a 70 by 20
pixel image covering R.A. $15^{h} 17^{m}-15^{h} 27^{m}$ and decl
$28^{\circ} 30'-29^{\circ} 18'$, in which each pixel brightness
represents a free parameter to be found. An iterative gradient
method was used to update the sky model and in turn calculate the
model scans and visibilities. For each pixel, the gradient was found
in order to reduce $\chi^2$ together with the constraint of
maximizing cross entropy of the sky model. To increase the speed of
the iterations the expected response in MITO and VSA data for each
trial pixel was pre-calculated and stored in a response matrix. This
algorithm was adapted from the version used in Dicker \ea (1999)
using the positive-negative cross-entropy method of Maisinger \ea
(1997). After 120 iterations, the solutions had converged and the
final sky models were convolved back to a common resolution of
$16'$. Because of the MITO beam-throw ($41'$) and the VSA primary
beam (2.$^{\circ}$0 FWHM), structures on degree scales and larger
are attenuated and with the present S/N are impossible to recover.
We are therefore limited in what we can say about the broader scale
profile of the signal.

A maximum likelihood analysis is then applied pixel by pixel on the
extracted maps in order to derive information about the presence of
primary and secondary anisotropies in our maps. Results are given in
section \ref{par:res}.

The analysis methodology and the codes have been tested using
different known signals embedded in both correlated and uncorrelated
noise and have yielded a satisfactory level of reproducibility of
the results up to a level of random noise of 15$\%$ compared to the
actual atmospheric noise (this set the minimum decorrelation
threshold of 85$\%$ between our channels). Different tests have been
carried out by changing the relative parameter describing the
signals as well as the quantities driving the fits and the
decorrelations, in order to optimize the latter for a more efficient
signal extraction.

\section{Contaminants}\label{par:cont}

Signal contamination due to galactic dust has been assessed through
the same simulation procedure described in section \ref{par:sim}.
The dust temperature distribution in the Corona Borealis region has
been extracted from the DIRBE-recalibrated IRAS 100 $\mu$m maps
presented in Schlegel {\it et al.} (1998), and extrapolated to the
MITO frequencies and bandwidths through the procedure described by
Finkbeiner {\it et al.} (1999), assuming their best-fit model for
the spectrum of emission. The observed region is quite contaminated
at high frequencies and presents structures at the resolution of our
interest: the flux estimates for the MITO bands and beam size lead
to thermodynamic temperatures of 8.2, 29, 62, and 208 $\mu$K at the
nominal coordinates of the H spot. Because of the adopted modulation
strategy and the smoothness of the dust distribution or the fact
that the mentioned structures are present over a diffuse signal, the
signals detected while drift scanning over the region are only
sensitive to the far smaller gradients at the angular scales of the
MITO chopping amplitude. Therefore, after simulating this observing
strategy over the extrapolated maps, the corresponding signals in
the observed direction turn out to be 0.2, 0.6, 1.3, and 4.2 $\mu$K
at most with a few percent variations due to sky rotation with
respect to the modulation axis. Even if small and substantially
negligible at the lowest frequencies, this contribution has been
accounted for and subtracted from the signal of each scan in the
data-set before performing the combined SZE-CMB extraction
procedures described above.

In order to quantify the possible contamination from point sources,
we have identified all radio sources in the region of observation in
the NVSS--1.4 GHz \cite{con98} and GB6--4.85 GHz \cite{gre96}
catalogs. From their fluxes we derived the spectral indexes that
were used to extrapolate the fluxes to the four MITO frequencies.
All the identified radio sources have predicted fluxes, in each of
the four channels, which have no imprint in our data. However, note
that we are limited by the sensitivity of the GB6 data, and there
could be some undetected sources with rising spectral index and a
high flux density at the MITO frequencies. To account for this, we
have obtained upper limits for the fluxes at the MITO frequencies of
all the identified sources in the NVSS catalog, by assigning a
maximum value of the rising spectral index of $\alpha=0.5$
($S\propto \nu^{\alpha}$). With this method we have found negligible
contamination to our observations too.

As we have described, constant and linear atmosphere emission has
been subtracted by the differential measurements and correlated
temporal and spatial fluctuations have been reduced by the
decorrelation procedure. Uncorrelated atmospheric noise is still
present in our data and will add to the detector noise. However, MEM
has been found to be efficient in extracting sky information from
uncorrelated noise. As anticipated in section \ref{par:noise}, a
source of residual atmospheric noise may be present in our data,
depending on the variation of PWV within the observational time.
This source of contamination is larger at higher frequencies and is
present as a common signal in the 214 and 272 GHz channels as a
residual of the imperfect decorrelation and of the variation of the
PWV content in the drift-scanning procedure.

\section{Results and discussion}\label{par:res}

The results of the analysis described in section \ref{par:anal} are
presented in this section. The maximum likelihood analysis on the
maps extracted from the MEM has been performed in two steps; by
extracting first common signals and then frequency-dependent signals
from the different observational channels. We notice that the
signals registered by the VSA and the 143 GHz MITO channel are
consistent within each other; in the same way, the 214 GHz and 272
GHz channels are consistent within each other. However, when
comparing the four channels at once to perform the extraction of the
CMB components characterizing our signals (primary anisotropy and
SZE), a third component is evident in the two high-frequency
channels, which does not allow the extraction of meaningful
information. A possible origin for the further component could be in
dust residuals, which have been carefully checked, or in atmospheric
residuals present in the 214 and 272 GHz channels due to variable
PWV, as suggested in section \ref{par:cont}.

In order to perform a meaningful analysis, we have therefore
calculated the signal separation at low frequency (\ie 33 and 143
GHz) and again at high frequency (\ie 214 and 272 GHz), and we use
the low-frequency determination of the primary anisotropy to infer
the third additional component. As a result, we find evidence of a
small signal common to all the four channels with a spectral
behavior typical of the SZE. This signal is restricted to an area of
approximately $15'$ in both right ascension and declination, with an
apparent shift towards lower right ascension with respect to the
primary anisotropy spot although the low S/N does not allow us to
fully describe the shape of the emitting region; the maximum signal
corresponds to the pixel at coordinates $R.A. \rm (J2000)=15^{h}
21^{m} 42^{s}$ and $\delta \rm (J2000)=28^{\circ} 50' 24''$ (in the
$6'$ resolution binned maps) with a maximum intensity of the
Comptonization parameter of $y=(7.8^{+5.3}_{-4.4})\times10^{-6}$
68\% CL. The signal observed by VSA is dominated by a primary
anisotropy with a small contribution from SZE. In a more
conservative attitude, with 95\% CL uncertainties, our observations
could put an upper limit of y$ < 1.8$ $10^{-5}$ to
diffuse/unknown-cluster SZ emission in the studied region.

In figure \ref{mappe} we show the maps of the observed sky region
derived from MEM for VSA and the 3 MITO channels (excluding the
353-GHz channel used for atmospheric subtraction) in terms of
thermodynamic temperatures. In the high-frequency channels we have
subtracted the third component found at those frequencies so that
only the CMB (primary and secondary effects) is present. In figure
\ref{mapperisult}, the maximum-likelihood-derived maps for the
primary anisotropy and for the SZ effect (Rayleigh-Jeans
thermodynamic temperatures) are given.

In figure \ref{mappay1dim} we present a cut at $\delta=28^{\circ}
50' 24''$ of the derived anisotropy maps extracted for primary and
secondary contribution.  A minimum is evident in the lower plot
corresponding to a RJ thermodynamic temperature of $\Delta
T_{SZ}=(-42^{+24}_{-29})\mu $K (RJ), with a primary anisotropy value
of $\Delta T_{ani}=(-128^{+21}_{-18}) \mu $K, corresponding to the
thermodynamic temperatures reported in figure \ref{bestfitmin}. The
primary anisotropy signal has been studied using the Monte Carlo
simulations method presented in Genova-Santos {\it et al.} (2005a):
the probability of finding such a spot in a CMB map is now increased
to 43$\%$. From the reported values, the fraction of the signal due
to SZE is (in the RJ regime) $f \equiv \Delta T_{SZ} / (\Delta
T_{SZ} + \Delta T_{ani}) = 0.25^{+0.21}_{-0.18}$. Note that these
values correspond to the maximum entropy sky reconstruction at the
resolution of the MITO experiment (\ie $16'$). In order to compare
with the VSA measurement at 33 GHz of $-230 \pm 23 \mu$K, we proceed
as follows. Using the same MEM code, we performed two different
reconstructions using only the VSA data: one at the MITO resolution
($16'$) and another one at the VSA resolution ($11'$). In these
maps, the H spot is observed with a peak temperature of $-174$ and
$-238 \mu$K, respectively. These numbers are fully consistent with
the inferred signal amplitude using MEM with MITO+VSA at the MITO
resolution, being also compatible with the observed amplitude of the
decrement in the VSA mosaiced map ($11'$ resolution) presented in
Genova-Santos {\it et al.} (2005a). These results further support
the robustness of MEM reconstruction.

We can now use the observed SZ signal to place constraints on the
physical parameters describing the gas. As in Genova-Santos {\it et
al.} (2005a), we shall consider two possibilities for the spatial
distribution of the gas.

First, given the angular extension of the feature in the y-parameter
map (it is not resolved by the $16'$ MITO beam), we can explore the
possibility of a cluster of galaxies as responsible for the
emission. We shall first estimate the amplitude of the SZ signal at
the VSA resolution. If we assume a point-like object, then at $11'$
we would expect around $-89 \mu$K (RJ temperature). This value is of
the same order as the expected SZ signal from the known cluster
members of Corona Borealis (see table 1 in Genova-Santos \ea 2005a).
However, it is unlikely that a cluster of galaxies at the redshift
of Corona Borealis would have been missed by previous optical
surveys. Nevertheless, the possibility of a cluster located at
larger distance that was spuriously aligned with the Corona Borealis
supercluster cannot be discarded.

On the other hand, we can also consider the case of WHIM being
distributed in a filament, aligned along the line of sight of the
supercluster with a physical dimension of some 40 Mpc. Using the
amplitude of the measured SZ component, we can set the constraints
presented in figure \ref{param} in the electron density-temperature
plane for such a filament. The shaded regions represent the
parameter space constrained from both the SZ observations and the
information extracted from the upper limit to the X-ray emission
(there is no evidence of emission in the R6 band of the ROSAT
XRT/PSPC All-Sky Survey, see Genova-Santos \ea 2005a). In figure
\ref{param} we also consider the +1 $\sigma$ and -1 $\sigma$ cases
for the fraction of the signal due to SZ. If we assume for the gas
temperature the typical values predicted for the WHIM (0.5-0.8 keV),
then we would expect baryon overdensities greater than 400-600 times
the mean baryon density in the local Universe, implying a baryonic
mass content ($\approx 3 \times 10^{14} M_\odot$) comparable to the
total baryonic mass of the cluster members of the supercluster.

\section{Conclusions}

We have presented the MITO observation of the Corona Borealis
supercluster. The analysis has been performed combining MITO results
with VSA observations of the same sky region. A MEM analysis has
allowed us to extract maps from our differential measurements. From
an analysis of these maps we find evidence of the presence of a
strong primary anisotropy signal with a faint secondary SZE signal
present in all four channels of the order of $-40\mu $K (RJ). The
detection is faint and it seems to be unresolved by the MITO beam.
Our results discard a non-Gaussian origin for the CMB spot
originally detected by VSA and favor the presence of hot-warm gas in
the intracluster region of Corona Borealis with a baryonic mass
content of $\approx 3 \times 10^{14} M_\odot$, even if the
hypothesis of an unknown unresolved cluster of galaxies cannot be
discarded. Higher sensitivity and angular resolution observations,
goals being already pursued both by superextended VSA (Taylor, 2003)
and MAD \cite{lamagna2002}, will be able to shed a definitive light
on the origin of the signal present in this region.

\acknowledgments The MITO team thanks the support of the Sezione
INAF in Torino. The Cosmology Group in Rome, a member of the
GEMINI-SZ Project, is supported by COFIN 2004 027755 and Ateneo
2004. Partial support is provided by "Azioni Integrate" IT 2196.
E.S.B. wishes to thank the CMBNET consortium for supporting his
work. We are grateful to J.R. Pardo for the ATM program. We would
like to a acknowledge Yoel Rephaeli and Mark Halpern for useful
comments to the paper.

All the work carried out in Rome and at the MITO observatory has
been guided, originated and performed under the leadership and
supervision of Professor Francesco Melchiorri who died prematurely
on 2005 July 28. His life, his work, and his teachings will long be
remembered.

\clearpage

\begin{figure}
\plotone{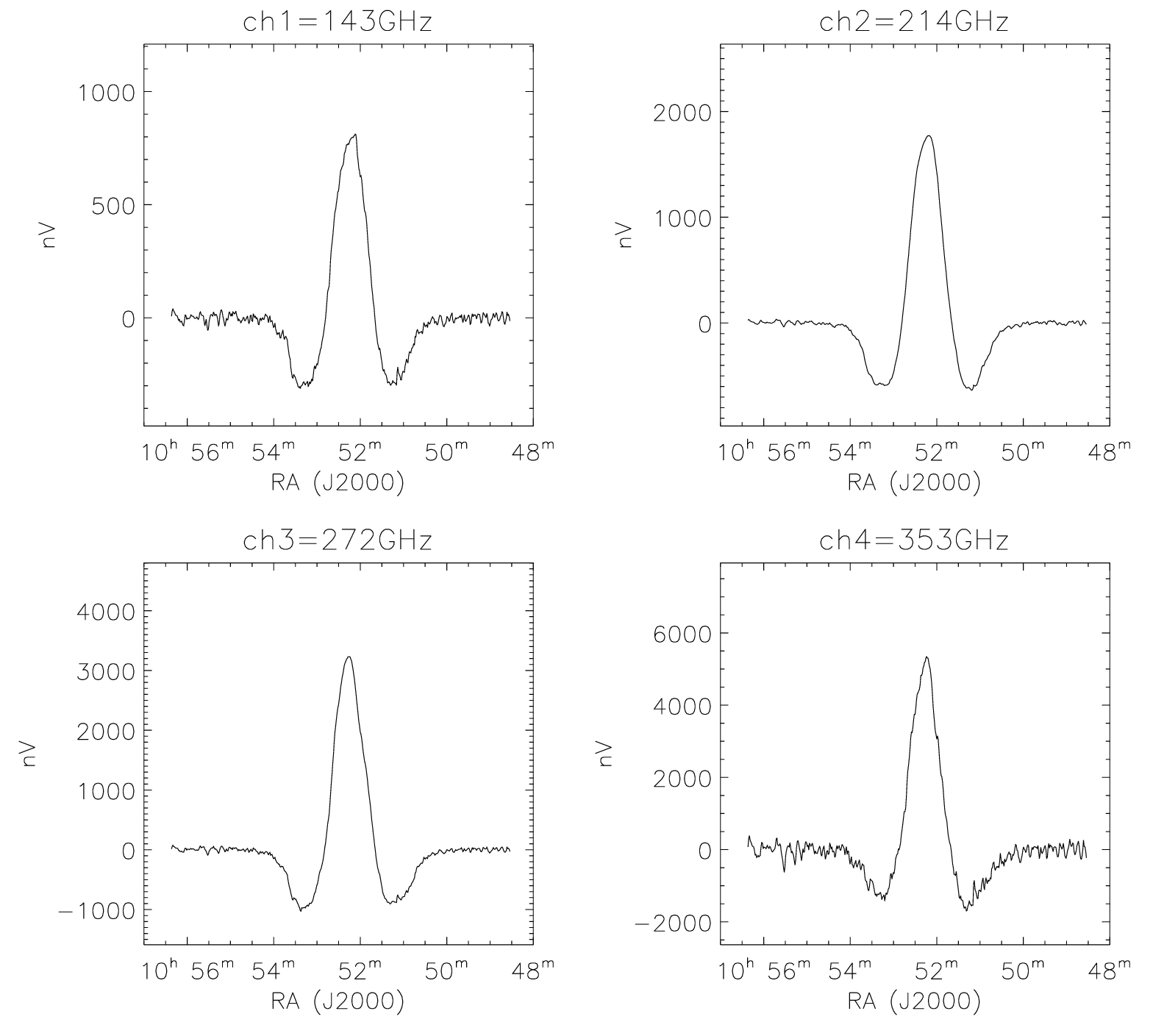} \caption{Signal detected observing Jupiter in the
four MITO channels. \label{jup}} \label{jup} \end{figure}

\clearpage

\begin{figure}
\plotone{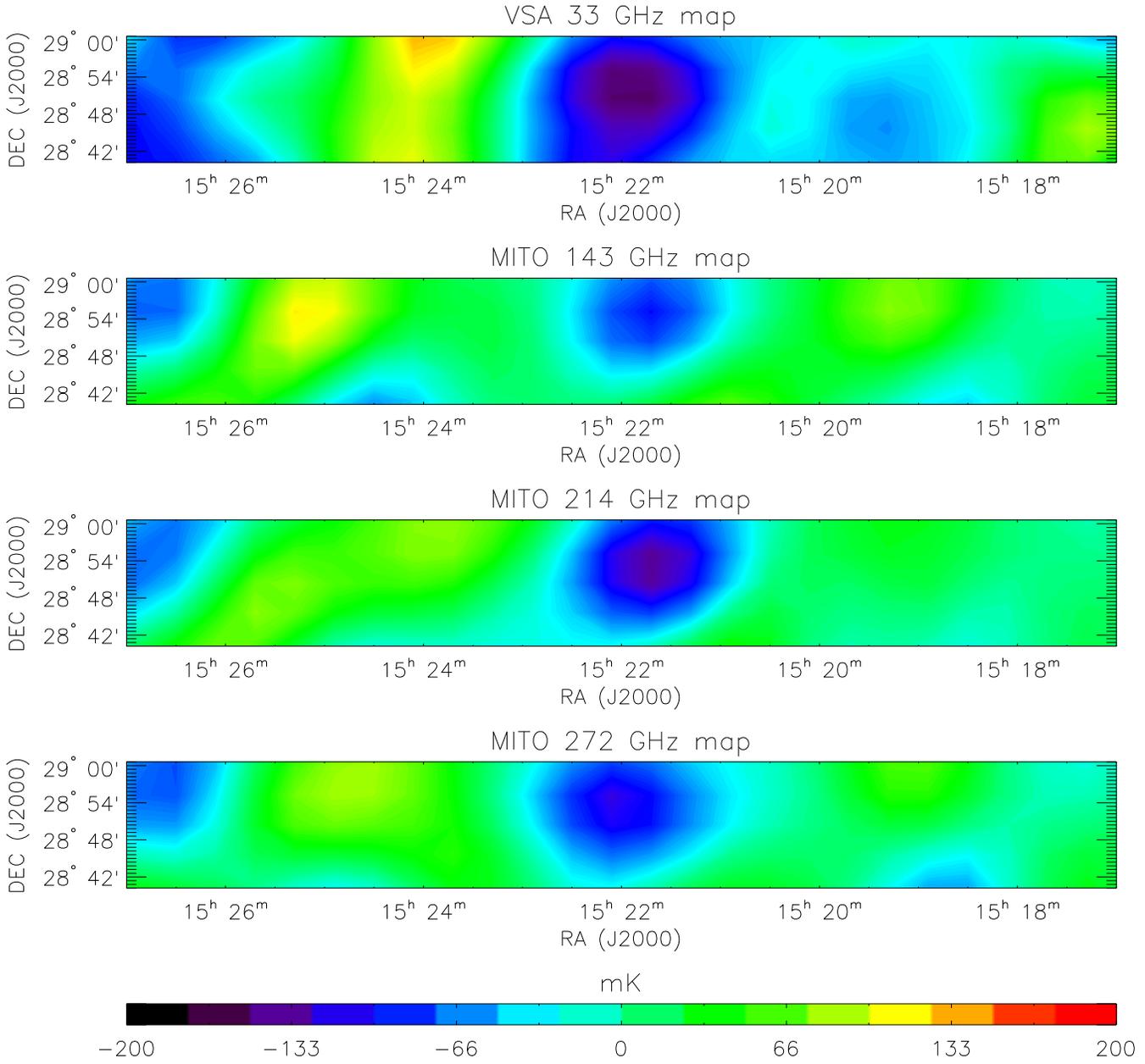} \caption{Strip of thermodynamic temperature
maps derived from the MEM for VSA and the three MITO channels.
\label{mappe}} \label{mappe} \end{figure}

\clearpage

\begin{figure}
\plotone{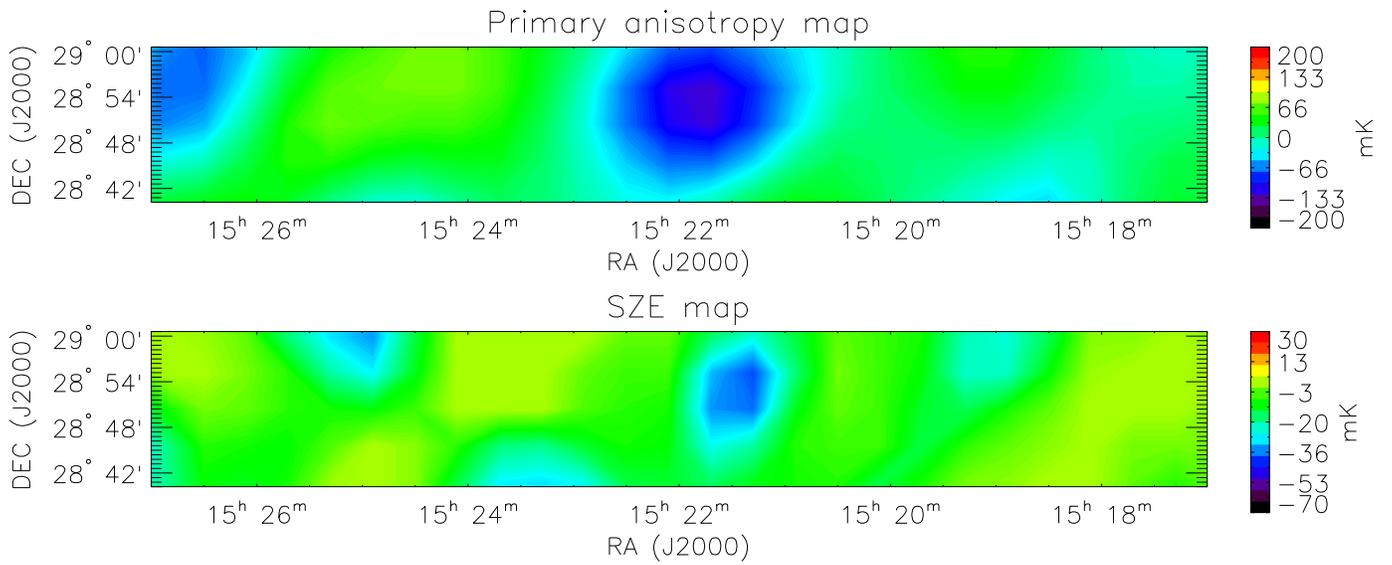} \caption{Primary anisotropy and SZE maps
derived by maximum likelihood. A spotis evident in the SZ map. This
seems to be not resolved by the MITO beam even though the low S/N
does not allow a complete description of it. \label{mapperisult}}
\label{mapperisult}
\end{figure}

\clearpage

\begin{figure}
\plotone{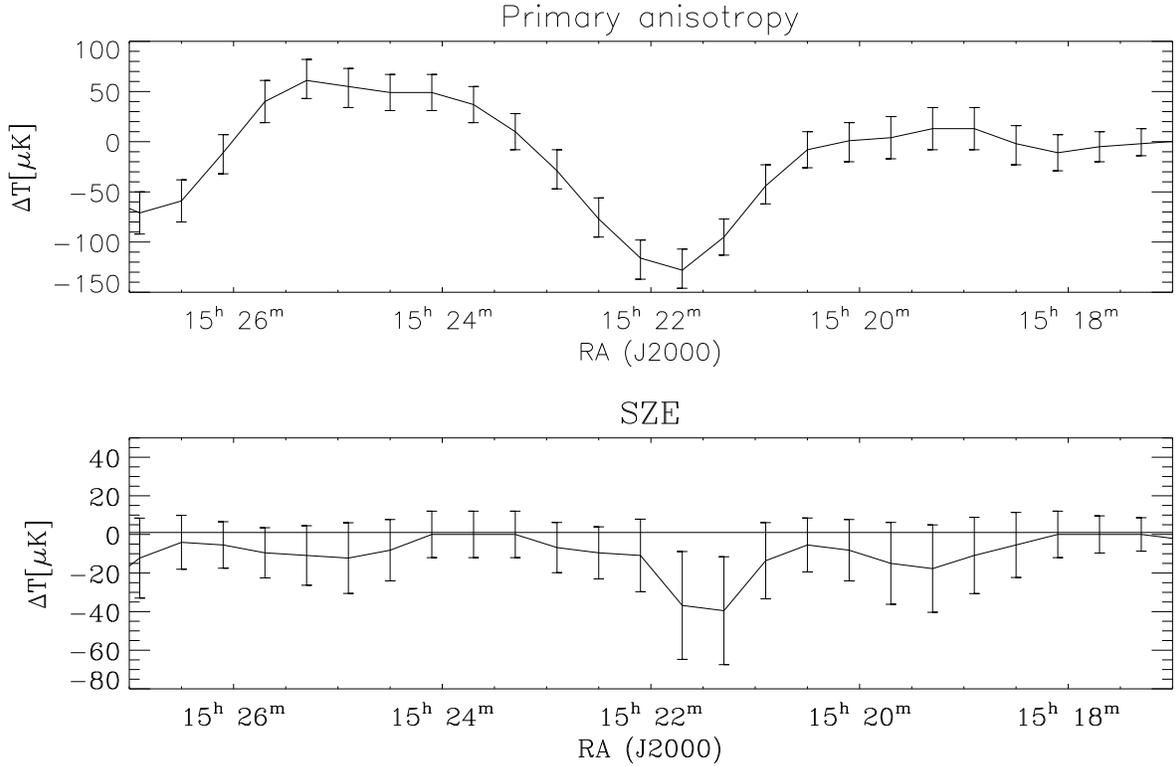} \caption{Primary and secondary SZE anisotropies
obtained from VSA and MITO observations at declination
$\delta=28^{\circ} 50' 24''$ with 1 $\sigma$errors. We have plotted
the thermodynamic temperature of the primary anisotropy and the
quantity $\Delta T=-2~y~T_{CMB} $ related to the SZE temperature in
the RJ part of the spectrum. There is also a clear minimum
corresponding to a maximum value of
$y=(7.8^{+5.3}_{-4.4})\times10^{-6}$ at $R.A. =15^{h} 21^{m}
42^{s}$. \label{mappay1dim}} \label{mappay1dim}
\end{figure}

\clearpage

\begin{figure}
\plotone{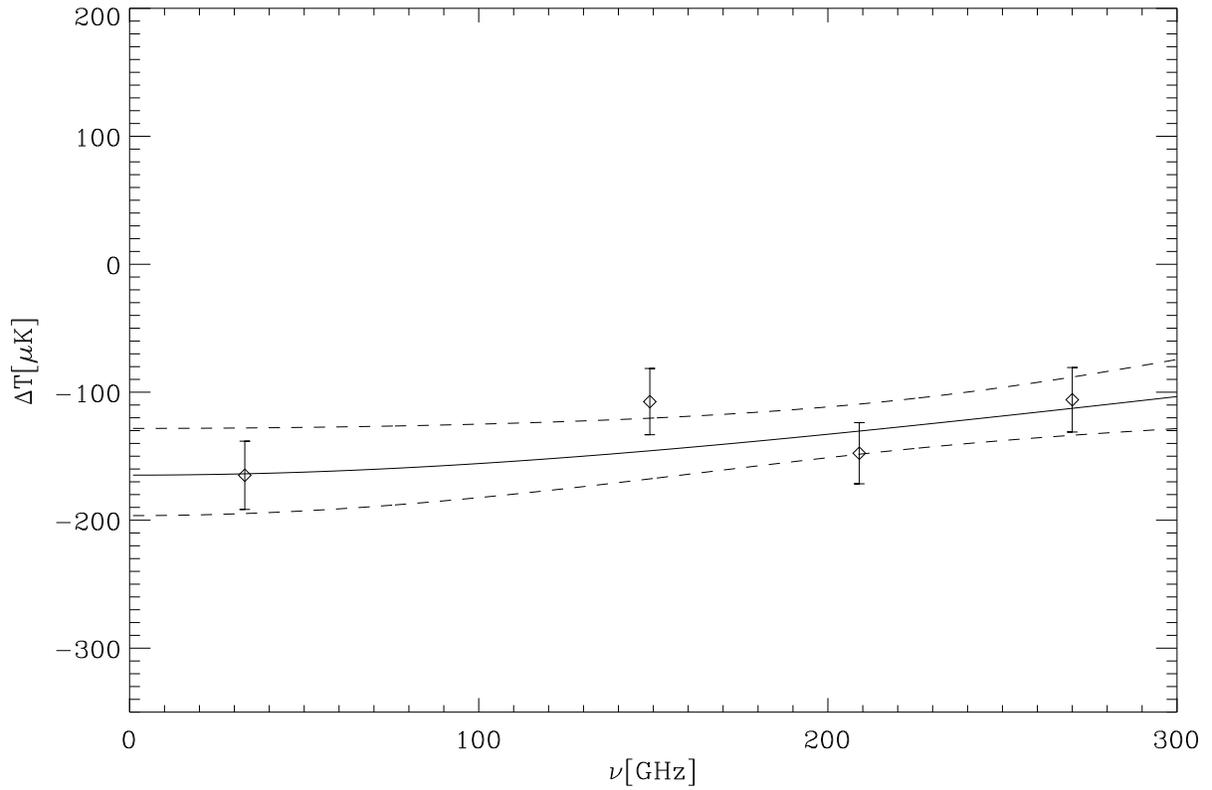} \caption{VSA and MITO detections plotted with the
spectral dependence of the resulting superposition of primary and
secondary anisotropy (solid line). Both the derivations have been
given with 1 $\sigma$ error (dashed line). This figure refers to the
maximum SZE signal observed in figure 5 at $R.A.=15^{h} 21^{m}
42^{s}$ and $\delta=28^{\circ} 50' 24''$. \label{bestfitmin}}
\label{bestfitmin}
\end{figure}

\clearpage

\begin{figure}
\plotone{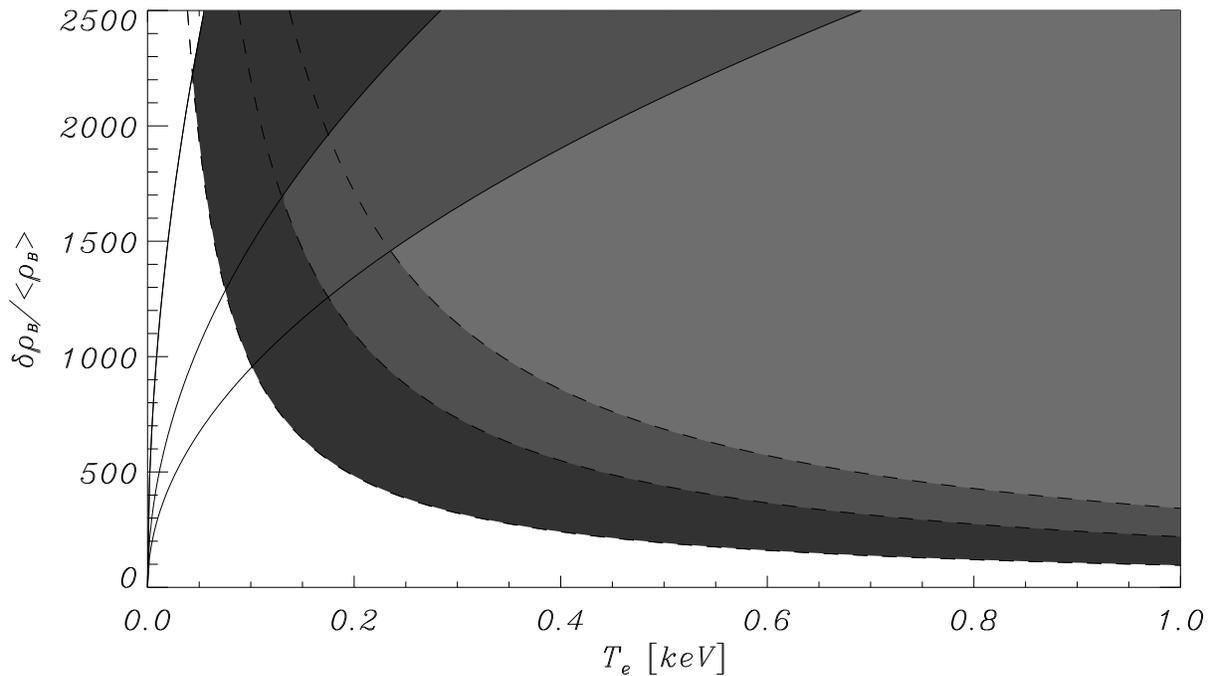} \caption{Constraints placed in the over-density
(with respect to the mean baryon density) and electron temperature
space if we set the SZ component to the nominal value inferred from
the analysis in this paper. The filled regions include the allowed
parameters combining the ROSAT R6 band observations (represented by
the solid line) and an SZE like that detected, obtained assuming a
maximum line of sight of 40 Mpc. Lighter and darker colors
correspond to -1 $\sigma$ and +1 $\sigma$ respectively.
\label{param}} \label{param}
\end{figure}

\clearpage

\begin{deluxetable}{ccccccc}
\footnotesize \tablewidth{0pt} \tablecaption{\it Main MITO
characteristics from the March 2004 Campaign. Central frequencies
$\nu$, bandwidths (BW), throughput (A$\Omega$), noise-equivalent
temperature (NET), average calibration factors (C), and atmospheric
transmissions ($\tau$) are included. \label{tab:char}}
\tablecolumns{6} \tablehead{ \colhead{Channel} &
\colhead{$\nu$(GHz)} & \colhead{$BW$(GHz)} &
\colhead{$A\Omega$(cm$^{2}$sr)} & \colhead{$NET$(mK$\sqrt{s}$)} &
\colhead{$<C>(\mu K/nV)$} & \colhead{$< \tau >(\%)$}} \startdata
     1 & 143 & 15 & 0.40$\pm$0.02 & 2.41$\pm$0.24 & 480$\pm$48 & 88 \\
     2 & 214 & 15 & 0.36$\pm$0.02 & 2.64$\pm$0.26 & 376$\pm$38 & 83 \\
     3 & 272 & 16 & 0.34$\pm$0.02 & 1.40$\pm$0.14 & 279$\pm$28 & 81 \\
     4 & 353 & 13 & 0.34$\pm$0.02 & 1.95$\pm$0.19 & 102$\pm$10 & 53 \\
\enddata
\end{deluxetable}

\clearpage

\begin{deluxetable}{lccc}
\footnotesize \tablewidth{0pt} \tablecaption{\it Average
Characteristics of the Decorrelated scans. We have included: the
average correlation factors for the first 3 MITO channels with the
353 GHz channel before and after the filtering procedure, the
coefficient of proportionality of the atmospheric fluctuations
between the various channels and the 353 GHz channel together with
the dispersion of these channels, the $rms$ before and after the
filtering procedure and the decorrelation. \label{tab:decor}}
\tablecolumns{4} \tablehead{ \colhead{Parameter} & \colhead{Ch.1} &
\colhead{Ch.2} & \colhead{Ch.3}} \startdata
     Correlation prefiltering (\%) & 74 & 73 & 81\\
     Correlation postfiltering (\%) & 95 & 95 & 96\\
     Atmospheric factors& 10.7 & 13.4 & 6.8\\
     Dispersion Atm. factors& 1.8 & 1.6 & 1.4\\
     $rms$ prefiltering (nV) & 9.7 & 7.7 & 15.3 \\
     $rms$ predecorrelation (nV) & 6.2 & 5.3 & 11.5 \\
     $rms$ postdecorrelation (nV) & 2.4 & 2.1 & 3.2 \\
    \enddata
\end{deluxetable}


\begin{thebibliography}{}

\bibitem[Banday \ea 1996]{ban96}
Banday, A.J., \ea 1996, {ApJ} {468} {L85}

\bibitem[Bardelli \ea 1996]{bard96}
Bardelli, S.,  \ea 1996, {A\&A} {305} {435}

\bibitem[Battistelli \ea 2002]{bat02} Battistelli, E.S., \ea 2002,  {ApJ} {580} {L101}

\bibitem[2003]{bat03} Battistelli, E.S., \ea 2003, {ApJ} {598} {L75}

\bibitem [Battistelli \ea 2005]{bat05}
Battistelli, E.,S., \ea 2005, {Proc. of the international school of
physics "Enrico Fermi" 6-16 July 2004} {IOS press} {Melchiorri and
Rephaeli eds} {379-387}

\bibitem[Bonamente 2003]{bon03} Bonamente, M., \ea 2003,
{ApJ} {585} {722}

\bibitem[Birkinshaw 1999]{birk99} Birkinshaw 1999,
{Physiscs Reports} {310} {97}

\bibitem[Carlstrom \ea 2002]{carl02}
Carlstrom, J.R., Holder, G.P., and Reese, E.D. 2002, {ARAA} {40}
{643}

\bibitem[Cen and Ostriker 1999]{cen99}
Cen, R., and Ostriker, J.P. 1999, {ApJ} {514} {1}

\bibitem[Condon \ea 1998]{con98} Condon, J.~J., \ea 1998, {AJ} {115} {1693}

\bibitem[Day \ea 1991]{day91}
Day, C.S.R., \ea 1991, {MNRAS} {252} {394}

\bibitem[De Gregori \ea 2006]{deg06}
De Gregori, S., \ea 2006  {in preparation}

\bibitem[De Petris \ea 1999]{depe99}
De Petris, M., \ea1999,   {New Astronomy} {43} {297}

\bibitem [De Petris \ea 2002]{depe02}
De Petris, M., \ea 2002,  {ApJ} {574} {L119}

\bibitem [2005a]{depe05a}
De Petris, M., \ea 2005a, {Proc. of the international school of
physics "Enrico Fermi" 6-16 July 2004} {IOS press} {Melchiorri and
Rephaeli eds} {345-354}

\bibitem [De Petris \ea 2005b]{depe05b}
De Petris, M., \ea 2005b, {EAS Publications Series} {14} {233-238}

\bibitem[Dicker \ea 1999]{dic99}
Dicker, S.R., \ea 1999, {MNRAS} {309} {750}

\bibitem[Kaastra \ea 2003]{kaa03}
Kaastra, S.R., \ea 2003, {A\&A} {397} {445}

\bibitem[Finoguenov \ea 2003]{fin03}
Finoguenov, A., 2003, {A\&A} {410} {777}

\bibitem[Finkbeiner \ea 1999]{fin99}
Finkbeiner, D.P., Davis, M., and Schlegel, D.J. 1999, {ApJ} {524}
{867}

\bibitem[Fukugita \ea 1998]{fuk98}
Fukucita, M., Hogan, C.J., and Peebles, P.J. 1998, {ApJ} {503} {518}

\bibitem[Genova-Santos \ea 2005a]{gen05a}
Genova-Santos, R., \ea 2005a,  {MNRAS} {363} {79}

\bibitem[2005b]{gen05b}
Genova-Santos, R., \ea 2005b, {Proc. of the international school of
physics "Enrico Fermi" 6-16 July 2004} {IOS press} {Melchiorri and
Rephaeli eds} {389-394}

\bibitem[Gregory \ea 1996]{gre96}
Gregory, P.C., \ea 1996, {ApJS} {103} {427}

\bibitem[Hern\'andez-Monteagudo \ea 2004a]{hmr04}
Hern\'andez-Monteagudo, C., and Rubi\~no-Mart\'{\i}n, J.A. 2004a,
{MNRAS} {347} {430}

\bibitem[2004b]{hmg04}
Hern\'andez-Monteagudo, C., Genova-Santos, R., and Atrio-Barandela,
F. 2004b, {ApJ}{613} {2} {L89-L92}

\bibitem[Lamagna \ea 2002]{lamagna2002} Lamagna, L., \ea 2002, {AIP} Conference Proc., 616,
92, M. De Petris and M. Gervasi eds.

\bibitem[Lieu \ea 2005]{lie05}
Lieu, F. 2005, {ApJ submitted} {astro-ph/0510160}

\bibitem[Maisinger \ea 1997]{mai9705}
Maisinger, K., Hobson, M.P., Lasenby, A.N., 1997, {MNRAS} {290}
{313}

\bibitem[Melchiorri and Olivo Melchiorri 2005]{melc05}
Melchiorri, F., and Olivo-Melchiorri, B. 2005, {Proc. of the
international school of physics "Enrico Fermi" 6-16 July 2004} {IOS
press} {Melchiorri and Rephaeli eds} {211-223}

\bibitem[Moreno 1998]{mor98}
Moreno, R. 1998, {PhD thesis, University of Paris VI}

\bibitem[Myers \ea 2004]{mye04}
Myers, A.D., \ea 2004, {MNRAS} {347} {L67}

\bibitem[Pardo \ea 2001]{par01}
Pardo, J.R., \ea 2001, {IEEE} {49/12} {1683-1694}

\bibitem[Persic \ea 1988]{pers88}
Persic, M., Rephaeli, Y., and Boldt, E. 1988, {ApJ} {327} {L1}

\bibitem[1990]{pers90}
Persic, M., \ea 1990, {ApJ} {364} {1}

\bibitem[Rauch \ea 1997]{rau97}
Rauch, M. \ea 1997, {ApJ} {489} {7}

\bibitem[Reaphaeli and Persic 1992]{rap92}
Reaphaeli, Y., and Persic, M. 1992, {MNRAS} {259} {613}

\bibitem[Rephaeli 1995]{reph95} Rephaeli, Y. 1995, {Ann. Rev. Astron. and Ap.} {33} {541}

\bibitem[Rubi\~no Mart\'{\i}n \ea 2003]{rub03}
Rubi{\~n}o-Mart{\'{\i}}n, J.A., \ea 2003, {MNRAS} {341,4} {1084}

\bibitem[Rubi\~no-Mart\'{\i}n \ea 2000]{rub00}
Rubi{\~n}o-Mart{\'{\i}}n, J.A., Atrio-Barandela, F., and
Hern\'andez-Monteagudo, C. 2000, {ApJ} {538}, {53}

\bibitem[Savini \ea 2003]{sav03}
Savini, G., \ea 2003,  {New Astronomy} {8-7} {727}

\bibitem[Schlegel \ea 1998]{sch98}
Schlegel, D.J., Finkbeiner, D.P., and Davis, M. 1998, {ApJ} {500}
{525}

\bibitem[Spergel \ea 2003]{spe03}
Spergel, D.N., \ea 2003, {ApJS} {148} {175}

\bibitem[Sunyaev and Zel'dovich 1972]{SunZel72}
Sunyaev, R.A. and Zel'dovich, Ya.B. 1972,  {Comm. Astrphys. Space
Phys.} {4} {173}

\bibitem[Taylor 2003]{tay03}
Taylor A., 2003, {New Astronomy Reviews}, {47} {11-12} {925-931}

\bibitem[Watson \ea 2003]{wat03}
Watson, R.A., \ea 2003, {MNRAS} {341} {1057}

\bibitem[Yoshida \ea 2005]{yos05}
Yoshida, N. \ea 2005, {ApJ} {618} {2} {L91-L94}

\bibitem[Zappacosta \ea 2005]{zap05}
Zappacosta, l. \ea 2005, {MNRAS} {357} {3} {929-936}

\end{thebibliography}
\end{document}